\documentclass[a4paper,11pt]{article}

\usepackage[a4paper,top=2cm,bottom=2cm,left=3cm,right=3cm,marginparwidth=1.75cm]{geometry}
\usepackage[utf8]{inputenc}
\usepackage[T1]{fontenc}
\usepackage[english]{babel}
\usepackage{amssymb,amsmath}
\usepackage{siunitx}
\PassOptionsToPackage{hyphens}{url}\usepackage{hyperref}
\usepackage{csquotes}
\usepackage{booktabs}
\usepackage{longtable}
\usepackage{adjustbox}
\usepackage{array}
\usepackage{graphicx}
\graphicspath{{figures/}}
\usepackage{url}
\usepackage{titlesec}
\usepackage{authblk}
\usepackage{xcolor}
\usepackage{caption}
\usepackage[protrusion=true,expansion=false]{microtype}
\usepackage{orcidlink}
\usepackage{hyperref}
\newcommand{\floatnotes}[1]{%
  \par\medskip
  \noindent
  \begin{minipage}{\linewidth}
    \footnotesize
    #1\par
  \end{minipage}%
}

\hypersetup{colorlinks=true,linkcolor=black,citecolor=black,urlcolor=blue}

\titleformat{\subsection}{\mdseries\itshape\large}{\thesubsection}{1em}{}
\titleformat{\subsubsection}{\mdseries\itshape\normalsize}{\thesubsubsection}{1em}{}

\usepackage[authoryear,round]{natbib}
\setcitestyle{aysep={,}}
\newcommand{\exfp}{\textit{Experience}}
\newcommand{\eduf}{\textit{Education}}

\title{\textbf{Mayoral Experience or Municipal Capacity? Negative-Outcome Evidence on Municipal Budget Execution in Peru}}

\author{Ciro Machacuay\,\orcidlink{0000-0002-0398-0302}%
\thanks{Email: \href{mailto:ciro.machacuay.m@uni.pe}{ciro.machacuay.m@uni.pe}}}

\author{Jaime Lincovil\,\orcidlink{0000-0003-4633-4033}%
\thanks{Email: \href{mailto:jlincovilc@uni.edu.pe}{jlincovilc@uni.edu.pe}}}

\author{Helder Rojas\,\orcidlink{0000-0002-6183-8929}%
\thanks{Email: \href{mailto:hrojas@uni.edu.pe}{hrojas@uni.edu.pe}}}

\affil{Universidad Nacional de Ingeniería, Lima, Perú}

\date{}
\begin{document}
\maketitle

\begin{abstract}

When experienced mayors govern better-performing municipalities, it is tempting to credit the leader. Yet municipalities with stronger administrative capacity may also be more likely to attract and elect experienced mayors, generating selection on pre-existing municipal capacity. We examine this identification problem using a balanced panel of 1,642 Peruvian district municipalities from 2015 to 2022. Within-municipality estimators and panel double machine learning recover a positive association between prior public-management experience and investment-budget execution, while formal education is substantially weaker. Negative-outcome controls reveal an important distinction. Across municipalities, mayoral human capital predicts GDP and HDI measured before the mayor took office, providing evidence of selection in cross-sectional comparisons. By contrast, the first-difference negative control aligned with the within-municipality design shows that changes in mayoral experience do not predict pre-determined district GDP, while they remain associated with budget execution. Sensitivity analysis and partial-identification bounds nevertheless do not support a clean causal interpretation of that within association. The results therefore separate two sources of variation often conflated in studies of political leadership and show how negative-outcome controls can sharpen the identification of leader effects.

\end{abstract}

\section{Introduction}
\label{sec:introduction}

Investment-budget execution is a central object of local public financial management: it is the rate at which authorised resources become accrued investment spending, and thus a proximate measure of whether municipal budgets translate into public goods \citep{worldbank2019,jacob2012,mef2024}. When that rate is regressed on the education or experience of the mayor, the coefficient is often read as evidence that leaders shape spending performance. The inference is intuitive, and it underwrites a familiar policy recommendation: raise the calibre of those who run cities, and cities will run better.

That reading sits within a broader tension in local-government research. An older view holds that municipal outcomes are set chiefly by structural constraints---fiscal, functional and socio-economic---leaving little room for whoever happens to occupy office \citep{peterson1981,tiebout1956}. A newer view holds that who governs matters: leaders' attributes leave measurable footprints on local policy and performance \citep{besley2011,avellaneda2009,avellaneda2022,wan2021}. In related work, \citet{carmeli2006} links the skills of municipal top-management teams to organisational performance, and \citet{debus2025} links partisan composition to local service provision. Public financial management research has examined fiscal autonomy, financial health and condition \citep{iacuzzi2025,zafragomez2009}, but less often subjects the mayoral coefficient on \emph{execution} to a falsification of selection.

This article argues that the leader-effect inference is fragile for a specific and diagnosable reason. Mayors are not randomly assigned to municipalities. In decentralised states with uneven administrative development, the districts able to attract, elect and sustain higher-human-capital mayors are frequently those that already possess stronger project banks, more stable technical teams and more professional investment-management routines. The coefficient on a mayor's curriculum vitae is then not an estimate of what the mayor contributes; it is a composite of the mayor's contribution and of the latent bureaucratic capacity of the jurisdiction that produced them. That is a claim about identification as much as about local politics, and it is testable.

We test it on a balanced panel of 1{,}642 Peruvian district municipalities over 2015--2022 (13{,}136 observations), spanning two complete mayoral terms and built from official budget-execution, electoral-CV and district-development records. We proceed in two movements. The first reproduces the conventional association, and does so generously: prior public-management experience is positively associated with investment-spending execution under a battery of within estimators and panel double-machine-learning (ATE $=0.0033$; 95\% CI $[0.0015, 0.0051]$), while formal education is weaker. A reader stopping there would conclude, with the existing literature, that mayoral human capital matters.

The second movement asks which part of that association survives a falsification the leadership literature almost never applies. We relate mayoral human capital to outcomes the mayor \emph{cannot have caused}---district GDP and HDI measured before 2015. Between districts, the test is decisive: mean experience predicts 2015 log GDP (0.094 SD, against 0.046 SD for the execution association). The test aligned with the within variation used by fixed effects is the first difference of human capital, not the level of the incoming cohort \citep{wooldridge2010}. That difference does not predict 2015 log GDP ($-0.0077$, SE $0.0110$), including among the 823 districts that actually change experience ($0.0087$, SE $0.0124$). The same step does predict execution. Cinelli--Hazlett robustness values are small (about 3.1\% of residual variance for experience and 2.3\% for education would nullify the corresponding within estimates). Benchmarking against log GDP after the within transformation is uninformative: that covariate is almost orthogonal to the 2019 step, so a 1$\times$ or 3$\times$ GDP adjustment leaves the point estimate unchanged. Mayor age is the least mute observed benchmark. Manski--Pepper bounds include zero; the reconstructable upper bound is $0.0097$, not the $0.0155$ reported in an earlier draft.

What follows is not a stronger causal claim about local leadership. Cross-sectional mayoral coefficients on investment-budget execution are joint estimates of the leader and of the district that selects the leader. The within-municipality association survives the aligned GDP falsification, is small, and is not a certified causal effect. Credential requirements for candidates therefore rest on shakier ground than the leadership literature's confidence suggests. The more useful object of study and of reform is the administrative apparatus---merit-oriented subnational civil service, continuity of technical teams, institutionalised project pipelines---as a hypothesis made more plausible, not as an effect identified here. Negative-outcome controls, which are cheap and portable, should become routine falsification tests for observational work on leader effects, as pre-trend tests became routine in difference-in-differences.

Peru is instructive because it combines heavy dependence on central transfers, sharp territorial variation in technical bureaucracies, and a ban on immediate mayoral re-election \citep{ley30305}\footnote{\citet{ley30305} prohibits the \emph{immediate} re-election of mayors and regional governors; a former mayor may return after sitting out a term. The ban applied from the 2018 municipal elections and therefore guarantees complete mayoral turnover between the two terms in our panel (2015--2018 and 2019--2022). The 2015--2018 cohort was elected in 2014 under the previous rules, so re-elected incumbents can appear in term~1 only. Mean experience falls from 3.84 years in term~1 to 2.08 in term~2: the rule creates identifying turnover and thins the stock of public-management experience.} that both creates the 2019 turnover we use and thins the experience stock. The between-district selection we document is the identification problem that any similarly decentralised, transfer-dependent setting should expect in \emph{cross-sectional} work; we do not treat Colombia, Mexico, Indonesia or the Philippines as estimated clones of this panel.

\noindent\textbf{Outline.}
The remainder of the paper is organized as follows. Section 2 develops the theoretical framework linking mayoral human capital, municipal bureaucratic capacity, and the selection of political leaders. Section 3 presents the data and empirical strategy, including within-municipality estimators, panel double-machine-learning, negative-outcome controls, sensitivity analysis, and partial identification. Section 4 reports the empirical results, moving from the conventional association between mayoral human capital and budget execution to the negative-outcome diagnostics that separate between-district selection from the within step. Section 5 discusses the implications for the interpretation of leader effects and for local public management, and Section 6 concludes.

\section{Literature review and theoretical framework}
\label{sec:literature}

The old view's canonical statement is \citeauthor{peterson1981}'s (\citeyear{peterson1981}) \emph{City Limits}: once socio-economic factors are accounted for, the balance of power among political actors has only modest policy consequences, because cities compete for mobile residents and capital and are therefore bound to pursue policies dictated by their structural position \citep{tiebout1956}. The newer view holds that local politics has nationalised and that leaders' attributes and ideologies now shape outcomes. Recent work suggests the two are complementary rather than rival: partisan footprints emerge where municipalities enjoy discretion and fiscal room, and recede where national regulation binds tightly \citep{breeman2015,gross2025,klindt2026}. \citet{anzia2021} argues that this question is best adjudicated not on spending aggregates but on the content of local \emph{service provision}, which is precisely what an execution ratio measures. The scope for leader effects is also conditioned by the structure of the local party system itself, which varies systematically with municipal size \citep{kjaer2010}, a consideration of direct relevance in a country of 1{,}642 districts ranging from a few hundred to several hundred thousand inhabitants.

Our argument sits at the hinge of that reconciliation, but adds a step the debate has largely skipped. If structure constrains what a municipality can do, structure also shapes \emph{who gets to try}. The two views are not merely complementary; they are entangled at the level of identification. We therefore place bureaucratic capacity at the centre of the theoretical framework. Comparative state-capacity research shows that development outcomes depend not only on elected leaders but also on the autonomy, merit orientation, and continuity of the administrative apparatus \citep{cingolani2013,dahlstrom2017,grindle2012}. In local governments, this means that mayoral education and experience can operate through two channels that are difficult to separate observationally: they may improve managerial decisions, but they may also proxy for the latent bureaucratic capacity of districts able to attract, select, and support higher-human-capital leaders.

\subsection*{Municipal fiscal performance and spending execution}

Spending execution is what translates budgetary resources into public goods and services \citep{worldbank2019,jacob2012}. In public financial management terms, the investment-execution ratio is a measure of \emph{budget credibility} on the expenditure side: the gap between what the modified institutional budget authorises and what is accrued in the fiscal year. Public financial management research has examined how fiscal arrangements relate to local financial health and condition \citep{iacuzzi2025,zafragomez2009,roberto2025}, but has concentrated on revenue-side indicators such as fiscal capacity and autonomy \citep{zafragomez2009,avellaneda2022,iacuzzi2025}. Investment-spending execution remains less explored, especially where implementation capacity---absorptive capacity for capital projects---is the binding constraint rather than headline fiscal space alone. We measure it as the ratio of accrued investment spending to the modified investment budget \citep{mef2024}. Over 2015--2022, mean execution across the 1{,}642 municipalities was 68.0\% (SD 22.2 percentage points).

\subsection*{Leader human capital and fiscal performance}

Whether that ratio moves with who holds office is a question of human capital as much as of fiscal design. Human-capital theory \citep{schultz1961,becker1964,mincer1958} treats skills and knowledge as a stock yielding returns: education supplies analytical capacity, experience supplies complementary tacit knowledge. \citet{heckman2000} adds that non-cognitive skills and institutional context are essential components of human capital, and these are largely unobservable in studies of political leaders. The implication matters for identification: because human capital is neither unidimensional nor fully observable, positive associations in conventional models must be treated as potentially fragile, which anticipates this study's emphasis on selection and sensitivity.

The evidence is mixed. \citet{freier2016} show that German voters prefer more-qualified mayors, though the effects on public finances are ambiguous. In Peru, available studies are mostly theses with geographically limited samples (Table~\ref{tab:related}). To our knowledge, no prior work has subjected the association between mayoral human capital and fiscal performance to a negative-outcome control \citep{lipsitch2010} capable of detecting selection on district capacity.

\begin{table}[htbp]
\centering
\small
\begin{adjustbox}{max width=\textwidth}
\begin{tabular}{llll}
\toprule
\textbf{Study} & \textbf{Sample} & \textbf{Design} & \textbf{Selection test}\\
\midrule
\citet{calleneyra2022} & Piura, first year of term & Cross-section & None\\
\citet{pomapalma2019} & Puno region, 2015--2018 & Cross-section & None\\
\citet{guevara2024} & 196 provincial municipalities, 2019--2022 & Department FE & None\\
This paper & 1{,}642 districts, 2015--2022 & TWFE / DML + NC & Between and $\Delta S$\\
\bottomrule
\end{tabular}
\end{adjustbox}
\caption{Peruvian studies of mayoral human capital and budget execution}
\label{tab:related}
\floatnotes{\textit{Notes:} All three prior studies read the mayor's CV as a determinant of execution without a negative-outcome control. \textit{Source:} Authors' own work.}
\end{table}

\subsection*{Bureaucratic capacity and the selection of leaders}

The fragile evidence just reviewed points beyond the mayor's CV. Administrative capacity is anchored in organisational human capital \citep{ingraham2003,otoole2009}. The mayor mobilises resources to attain public objectives; their human capital shapes their capacity for strategic articulation, coordination and problem-solving \citep{boyne2003}. Education contributes codified knowledge (regulations, project appraisal); experience supplies tacit knowledge (navigating the bureaucracy, informal relationships). \citet{otoole1999} propose that leaders enhance the contribution of their staff, generating a multiplier on organisational human capital \citep{melton2017}; \citet{carmeli2006} showed that it is the skills of the municipal top-management \emph{team}, not the elected leader alone, that predict organisational outcomes, an early indication that the apparatus carries independent weight.

That multiplier logic, however, cuts both ways, and this is the pivot of our argument. If a district's technical team is what converts a mayor's skill into executed investment, then the team is not merely a mediator of the mayor's effect; it is also a \emph{cause of who becomes mayor}. Districts with consolidated bureaucracies attract more credentialed candidates, offer more plausible platforms, and sustain the political careers from which experienced mayors emerge. This reading connects with the literature on bureaucratic capacity in Latin America, which stresses that individual leadership operates within, and is frequently dominated by, the technical capacity of the administrative apparatus \citep{idb2014,caf2015}, and with the diagnosis of Peru's own civil-service authority, the Autoridad Nacional del Servicio Civil (SERVIR), whose reform agenda targets precisely the professionalisation of subnational technical teams \citep{servir2021}. It is consistent with comparative work on state capacity and civil-service reform: bureaucratic autonomy and meritocratic recruitment condition whether political leadership translates into implementation capacity \citep{cingolani2013,dahlstrom2017,grindle2012}.

The implication for identification is direct. If mayoral human capital is correlated with the unobserved technical capacity of the municipal bureaucracy, because more-capable districts both elect more-credentialed mayors and sustain stronger project banks, then the leader-attribute coefficients estimated in the prior literature---especially in cross-section---partly capture latent bureaucratic capacity rather than a leadership effect. District fixed effects remove \emph{time-invariant} district capacity. They do not automatically remove capacity that evolves and co-moves with the incoming mayor. Whether that residual channel is present is an empirical question for a negative-outcome control aligned with the 2019 step, not a premise (Section~\ref{sec:findings}).

\subsection*{Hypotheses}

The framework implies four testable statements. Formal education should be positively associated with investment-spending execution in conventional models, but the association is expected to be highly sensitive to unobserved confounding (H1). Prior public-management experience should also be positively associated with execution, more robustly than education across within estimators, though still vulnerable to moderate confounding (H2). The COVID-19 period may alter the return to experience relative to education if experiential knowledge matters more under institutional stress (H3); we treat that claim as exploratory. Finally, between-district human-capital--execution associations should be contaminated by selection on pre-determined district conditions, while the within-district association should be subjected to---and can survive---a first-difference negative control aligned with the 2019 step (H4).

\section{Methodology}
\label{sec:method}

\subsection*{Data and sample}

The empirical design requires a national panel in which mayoral attributes and budget execution can be linked for complete terms. Starting from the universe of Peruvian \emph{district} municipalities (about 1{,}696 units in the period covered by our source catalogues), we retain the 1{,}642 districts for which investment-execution series, mayoral CVs and district covariates can be linked for every year 2015--2022, yielding a balanced panel of 13{,}136 observations. Roughly 54 districts (about 3\% of the district universe) drop out because at least one source is incomplete over the full window; the public analysis file contains only the balanced panel, so we cannot profile excluded districts on observables. If incomplete reporting correlates with weaker administrative capacity---as is plausible for Sistema Integrado de Administraci\'on Financiera (SIAF) non-reporters---balancing truncates the lower tail of capacity and would, if anything, \emph{understate} the selection channel this paper documents. Sources are official: municipal budget execution from the MEF's \emph{Consulta Amigable} platform \citep{mef2024}; mayoral biographical and professional attributes from elected-candidate CVs published by the JNE \citep{jne2022}; contextual variables from INEI \citep{inei2018} and the UNDP district Human Development Index \citep{undp2019}; and SUNEDU \citep{sunedu2022} and Ministry of Education \citep{minedu2022} records on the presence of universities and institutes. Separately, a pre-specified influence-screening rule (Cook's $D > 4/n$ and leverage above three times the mean, in at least half of the main specifications) flagged 607 observations (4.6\%) within the balanced panel. Main tables use the unscreened balanced panel.

\subsection*{Variables}

The outcome is the investment-budget execution ratio: accrued investment spending divided by the modified institutional budget (PIM), expressed as a proportion in $[0, 1]$. (In the replication files this variable is coded \texttt{EPGP}.) The measure is bounded, with appreciable mass at both limits. We therefore distinguish two reporting roles. A fractional logit is the bounded-outcome \emph{complement} (it accommodates observations at exactly 0 and 1); we do not treat the pooled fractional logit as a preferred specification, because it is identified from between-district variation that the negative controls show to be selected. The \emph{headline linear} association reported in the abstract and in the selection discussion is the panel double-machine-learning ATE after two-way demeaning (LassoCV nuisances), which matches the TWFE point estimate and is the natural counterpart to the within-aligned negative controls. Linear FE/TWFE, first differences and related estimators complete the battery (Table~\ref{tab:main}). More consequentially, the denominator is not exogenous. The PIM is modified during the fiscal year, and municipalities that anticipate poor execution may reduce it, mechanically inflating the ratio. Initial institutional-budget (PIA) amounts are not in the analysis file, so we do not report a PIA-denominator specification.

The two treatments of interest are years of formal education and years of prior public-management experience, both from JNE CVs. In the replication files these are coded \texttt{A\_EDUF} and \texttt{A\_EXFP}, respectively; a codebook maps all analysis labels to dataset names. Six observations with negative experience values, a coding convention of the electoral authority, were recoded to zero; results are invariant to this decision and to a $\log(1+\text{experience})$ specification.\footnote{Six observations carried negative experience values, a coding convention of the electoral authority; these were recoded to zero. Results are invariant to this decision and to a $\log(1+\text{experience})$ specification.} We control for the mayor's age, gender and professional category; an indicator of complaints or sentences on the electoral record; urban/rural status and natural region; log district GDP and district HDI; and provincial presence of universities and higher institutes as proxies for local institutional-technical capacity. Because execution ratios can be mechanically correlated with budget size (small budgets tend to exhibit higher execution rates), we also include the log per-capita modified investment budget and transfer-intensity categories, and report robustness to their exclusion (Section~\ref{sec:findings}).

\subsection*{Estimation strategy}

The hypotheses cannot be settled by a single estimator. We therefore use a battery of partially non-nested approaches---district and two-way fixed effects, panel double-machine-learning with cluster-level cross-fitting, a dose-response design for the pandemic, negative-outcome controls, Manski--Pepper bounds, and Cinelli--Hazlett/Oster sensitivity metrics with multiple-comparison correction. All estimates reported below are in the main text.

\subsubsection*{Model specification}
\label{sec:model}

Let $i$ index district municipalities and $t$ years. The outcome $Y_{it}$ is the investment-budget execution ratio (replication code \texttt{EPGP}). The treatments of interest are years of prior public-management experience $S_{it}$ (\texttt{A\_EXFP}) and years of formal education $E_{it}$ (\texttt{A\_EDUF}); the vector $X_{it}$ collects the controls listed above. The two-way fixed-effects (TWFE) specification that anchors the battery is
\begin{equation}
\label{eq:twfe}
Y_{it}=\alpha_i+\delta_t+\beta_S S_{it}+\beta_E E_{it}+X_{it}'\gamma+\varepsilon_{it},
\end{equation}
where $\alpha_i$ is a district fixed effect and $\delta_t$ a year fixed effect. Standard errors are clustered by district. Because $S_{it}$ and $E_{it}$ are constant within each four-year term, each district contributes only two treatment values over 2015--2022; the within variation in~\eqref{eq:twfe} is a single step at the 2019 transition.

First differences of~\eqref{eq:twfe} eliminate $\alpha_i$ and, under homogeneous effects and strict exogeneity, recover the same $\beta_S$ and $\beta_E$ as entity fixed effects. In this design the two need not coincide: the 2018--2019 cohort step loads heavily on the first-difference estimator, and serial correlation or term-coincident shocks can drive a wedge between first differences and TWFE (Section~\ref{sec:findings}).

For the linear DML check \citep{chernozhukov2018}, write the partially linear model after two-way demeaning as
\begin{equation}
\label{eq:dml}
Y_{it}- \ell_0(X_{it})=\theta\bigl(S_{it}-m_0(X_{it})\bigr)+\zeta_{it},
\end{equation}
with nuisance functions $\ell_0$ and $m_0$ estimated by cross-validated Lasso (or ridge as a check). We two-way demean $Y$ and $S$ by district and year \emph{before} cross-fitting so that residualised variables retain only within variation, form folds at the district level, and cluster standard errors by district. DML relaxes functional-form assumptions on the controls; it does not relax unconfoundedness. The headline DML ATE in the text is the LassoCV specification; the ridge check is reported alongside other estimators in Table~\ref{tab:main}. Fractional logit complements~\eqref{eq:twfe} as a bounded-outcome check; the panel-DML ATE is the headline linear summary for comparison with the negative-control diagnostics below.

\subsubsection*{Fixed effects (FE and TWFE)}
We begin with the within estimator in~\eqref{eq:twfe}. District fixed effects remove selection bias attributable to time-invariant district characteristics; year fixed effects absorb aggregate shocks common to all districts. Under heterogeneous treatment effects, TWFE estimators may incur negative weighting \citep{goodmanbacon2021,dechaisemartin2020}.

A structural feature of the design matters more than the textbook setup suggests. Mayoral education and experience are fixed at election and therefore constant \emph{within} each four-year term. Each district contributes \textbf{two} values of the treatment over the eight-year panel, not eight: the within variation used by the fixed-effects estimators is a single step, at the 2019 transition. Power is lower than the 13{,}136 observations suggest, and any district-specific shock that coincides with the change of term---including the pandemic one year later---is a candidate confounder that year fixed effects alone cannot absorb. We return to this in Sections~\ref{sec:findings} and~\ref{sec:discussion}.

\subsubsection*{Instrumental variables and their honest discarding}
An instrument would be attractive if valid. We explored instrumenting education with its lags and the lagged district HDI. Although the first-stage $F$-statistic is large, Hansen's overidentification test rejects joint instrument validity ($p<0.001$), and a plausibly-exogenous analysis \citep{conley2012} detects no significant effect at any level of exclusion violation. We discard the IV strategy and report instead partial-identification bounds \citep{manski2000}, which require no exclusion restriction. Reporting the failure is preferable to sustaining an identification the data do not support.

\subsubsection*{Double machine learning}
As a semiparametric check on the parametric within estimates, we implement~\eqref{eq:dml} with LassoCV nuisance models \citep{chernozhukov2018}. Agreement with the parametric estimate speaks to linearity, not to selection on unobservables---the question addressed in Section~\ref{sec:findings}.

\subsubsection*{The pandemic as a common-timing shock with continuous intensity}
The pandemic specification must match the shock. Unlike staggered-adoption settings \citep{sun2021,callaway2021,borusyak2024}, COVID-19 hits all municipalities in 2020 and interacts with a continuous mayoral trait (experience). The estimand is the dose-response of the experience--execution relationship before and after the shock \citep{callaway2024}. Our reference specification is a triple-difference (DDD) estimator; staggered-adoption estimators appear only as diagnostics.

A further complication, which we treat as a limitation rather than as solved, is that the mayoral cohort changes in 2019 and the pandemic arrives in 2020. The post-2020 distribution of experience differs from the pre-2020 distribution partly \emph{because the mayors are different people}. We therefore also report a specification restricted to the second term (2019--2022), within which the mayor is fixed.

\subsubsection*{Heterogeneity and negative controls}
Beyond average associations, we estimate effect heterogeneity with honest causal forests \citep{athey2019} and use negative-outcome controls \citep{lipsitch2010} as the direct diagnostic of residual selection. Alongside lagged and structural outcomes (lagged HDI, district GDP), we use strictly pre-determined outcomes measured \emph{before 2015}, prior to any mayoral term in the panel. Write $W_d$ for a pre-determined district outcome (log 2015 district GDP or 2015 HDI) and let $\bar S_d$, $\bar E_d$ denote district means of experience and education over the panel. The \emph{between} negative control is
\begin{equation}
\label{eq:nc-between}
\bar W_d=\rho_0+\rho_1 \bar S_d+\rho_2 \bar E_d+\text{region FE}+u_d.
\end{equation}
Equation~\eqref{eq:nc-between} detects between-district selection---the correlation that district fixed effects remove by construction. Because $S$ is constant within each term, the variation that identifies FE, TWFE and DML is the single 2019 step $\Delta S_d=S_d^{(2)}-S_d^{(1)}$, which is equivalent to the first-difference estimator under a one-time change of treatment \citep{wooldridge2010}. The \emph{aligned} negative control is therefore
\begin{equation}
\label{eq:nc-fd}
W_d^{2015}=\alpha_0+\alpha_1\Delta S_d+\alpha_2\Delta E_d+\text{region FE}+e_d,
\end{equation}
with decisive tests $H_0:\alpha_1=0$ and $H_0:\alpha_2=0$. We also report the levels specification
\begin{equation}
\label{eq:nc-within}
W_d^{2015}=\lambda_0+\lambda_1 S_d^{(2)}+\lambda_2 S_d^{(1)}+\lambda_3 E_d^{(2)}+\lambda_4 E_d^{(1)}+\text{region FE}+v_d,
\end{equation}
which detects selection on the \emph{level} of the incoming cohort. A pattern $\lambda_1>0$ together with $\lambda_2>0$ of similar size is the signature of between-district selection and is exactly what produces $\alpha_1\approx 0$. About half of districts have $\Delta S_d=0$, so we also report~\eqref{eq:nc-fd} on changers only, with the associated minimum detectable effect.

The negative-control outcome (log district GDP) and the main outcome (a proportion in $[0, 1]$) are measured on different scales, so raw coefficients are not comparable. We express both in standard-deviation units before comparing channels. For the between negative control, the treatment is the district mean of experience, so standardisation must use the standard deviation of that mean (1.920 in the reported negative-control standardisation), not the observation-level standard deviation (approximately 3.04); using the latter inflates the selection association. SD($\bar S_d$) $=1.920$ is the empirical standard deviation of the collapsed district-mean experience variable used in the district-level negative-control regression; the between SD of $1.953$ reported in Table~\ref{tab:variation} is the between component of the panel variance decomposition and is not the standardiser used for the negative-control comparison. Likewise, the 2015 log-GDP standard deviation used for the between control ($1.615$) is the cross-section dispersion in that single year and is smaller than the full-panel log-GDP standard deviation in Table~\ref{tab:desc} ($1.957$), because the panel adds within-district time-series variation.

\subsubsection*{Proportionality and the interpretation of selection bias}
\label{sec:proportionality}
Let $U_d$ denote latent district administrative capacity. Under the within transformation (double dots), the probability limit of the experience coefficient in~\eqref{eq:twfe} decomposes as
\begin{equation}
\label{eq:bias}
\mathrm{plim}\,\hat\beta_S=\beta_S+\frac{\mathrm{Cov}(\ddot S,\ddot U)}{\mathrm{Var}(\ddot S)}\,\frac{\partial Y}{\partial U}.
\end{equation}
A negative-outcome control such as~\eqref{eq:nc-between}--\eqref{eq:nc-within} informs the first factor---selection of human capital on $U$---and the reduced-form map from $U$ to $W$, but not $\partial Y/\partial U$. Passing from ``the negative control is predicted'' to ``a non-trivial share of $\hat\beta_S=0.0033$ is the district'' therefore requires an explicit \emph{calibration} (proportionality) assumption,
\begin{equation}
\label{eq:kappa}
\frac{\partial Y}{\partial U}=\kappa\,\frac{\partial W}{\partial U},\qquad \kappa\neq 0,
\end{equation}
for some scale factor $\kappa$ that links the capacity--execution channel to the capacity--negative-control channel. Without~\eqref{eq:kappa}, the negative control is a diagnostic of selection, not a bias-corrected average treatment effect. We state~\eqref{eq:kappa} as a maintained calibration condition, report the diagnostic (including the first-difference specification~\eqref{eq:nc-fd} in Section~\ref{sec:findings}), and do not claim a debiased ATE.

\subsubsection*{Sensitivity, inference, and multiple comparisons}
Finally, we report Cinelli--Hazlett robustness values \citep{cinelli2020} with covariate benchmarking (how strong a confounder would need to be relative to log district GDP after the within transformation) and Oster's $\delta$ \citep{oster2019}, together with region-year fixed effects, Driscoll--Kraay standard errors, and two-way clustering by district and year. We do not invoke a fixed ``conventional'' robustness-value percentage threshold; \citet{cinelli2020} recommend benchmarking against observed covariates. Given the number of heterogeneity tests, we control the false-discovery rate (FDR) via Benjamini--Hochberg and a Romano--Wolf stepdown procedure.

\section{Findings}
\label{sec:findings}

\subsection*{Descriptive statistics and the support for fixed effects}

\begin{figure}[htbp]
\centering
\includegraphics[width=\textwidth]{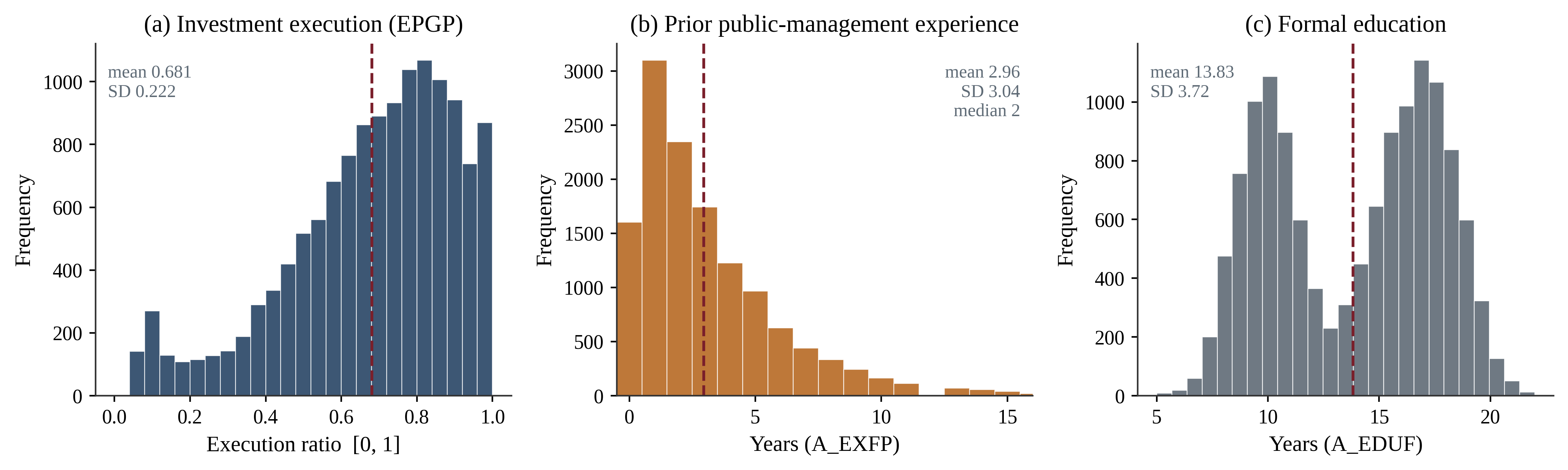}
\caption{Distribution of the panel's main variables}
\label{fig:dist}
\vspace{2pt}
\floatnotes{\textit{Notes:} Execution concentrates towards the upper end of the $[0, 1]$ interval, with appreciable mass at the bounds, motivating the complementary use of fractional-response models. Prior experience is strongly right-skewed (median of 2 years). \textit{Source:} Authors' own work.}
\end{figure}

Table~\ref{tab:desc} summarises the analysis sample, and Figure~\ref{fig:dist} plots the distributions of the main variables. Mean execution is 0.680 (SD 0.222), distributed across the full $[0, 1]$ range with mass near the upper bound. Mayors average 13.83 years of education (SD 3.72) and only 2.97 years of prior public-management experience (SD 3.04; median 2). About 3.6\% of observations are female mayors; 68\% of district--years are rural on the source zone label (\texttt{Zona}; the 32\% figure in an earlier draft was the urban share). The Spearman correlation between education and experience is just 0.04, so collinearity is not a concern; about 18\% of non-missing experience observations are exactly zero, which also cautions against treating the linear years specification as a complete description of the experience margin. The correlation between execution and human capital is modest (experience: 0.06; education: 0.01), anticipating small effect magnitudes. Modified institutional budget (PIM) levels in soles are not in the public analysis file (only the execution ratio and related controls), so absolute-budget magnitudes in the discussion remain illustrative.

\begin{table}[htbp]
\centering
\small
\begin{adjustbox}{max width=\textwidth}
\begin{tabular}{lrrrrrrrr}
\toprule
\textbf{Variable} & \textbf{$N$} & \textbf{Mean} & \textbf{SD} & \textbf{Min} & \textbf{p25} & \textbf{Med.} & \textbf{p75} & \textbf{Max}\\
\midrule
Investment execution ratio & 13{,}136 & 0.680 & 0.222 & 0.000 & 0.534 & 0.723 & 0.858 & 1.000\\
Mayoral experience (years) & 13{,}025 & 2.966 & 3.035 & 0.000 & 1.000 & 2.000 & 4.000 & 22.0\\
Mayoral education (years) & 13{,}136 & 13.83 & 3.722 & 0.000 & 11.00 & 14.00 & 16.00 & 27.0\\
Mayor age (years) & 13{,}028 & 50.42 & 9.669 & 21.00 & 44.00 & 50.00 & 57.00 & 100.0\\
Female mayor (share) & 13{,}136 & 0.036 & 0.187 & 0.000 & 0.000 & 0.000 & 0.000 & 1.000\\
Ethical-record complaints (count) & 13{,}136 & 0.099 & 0.460 & 0.000 & 0.000 & 0.000 & 0.000 & 6.000\\
Rural district (share) & 13{,}136 & 0.680 & 0.466 & 0.000 & 0.000 & 1.000 & 1.000 & 1.000\\
District HDI & 13{,}136 & 0.382 & 0.135 & 0.067 & 0.279 & 0.356 & 0.472 & 0.864\\
Log district GDP & 13{,}136 & 15.46 & 1.957 & 9.506 & 14.32 & 15.36 & 16.47 & 23.62\\
Universities in province (count) & 13{,}136 & 1.655 & 5.916 & 0.000 & 0.000 & 0.000 & 1.000 & 41.0\\
Higher institutes in province (count) & 13{,}136 & 32.55 & 98.12 & 0.000 & 4.000 & 8.000 & 21.00 & 606.0\\
\bottomrule
\end{tabular}
\end{adjustbox}
\caption{Descriptive statistics, analysis sample}
\label{tab:desc}
\floatnotes{\textit{Notes:} Balanced panel of 1{,}642 district municipalities, 2015--2022. Experience recodes six negative electoral-authority codes to zero (Section~\ref{sec:method}). Female mayor is the minority gender code in the JNE file (share $0.036$). Rural is the source zone label (\texttt{Zona}; share $0.680$). An earlier draft inverted that indicator and reported the urban share as rural. Log district GDP is $\ln(\mathrm{PBI\_DIST})$. The university and higher-institute counts are assigned at provincial (not district) scale: every district of Lima province carries the same 606 institutes, and almost all districts have a positive institute count (2015 share $0.997$). The maximum mayor age of 100 is a single observation (Quinuabamba, 2022). PIM levels in soles are not in the analysis file. \textit{Source:} Authors' own work, based on MEF, JNE, INEI and UNDP data.}
\end{table}

A precondition for the within estimators that anchor our identification is that the treatment variables carry meaningful within-district variation. Table~\ref{tab:variation} confirms that they do. The within standard deviation is 76.6\% of the total standard deviation for experience and 67.8\% for education; in variance terms, within variation accounts for 58.7\% and 46.0\% of total variation respectively. This variation is institutionally induced rather than incidental: the constitutional ban on immediate re-election \citep{ley30305}, applied from the 2018 elections, forces mayoral turnover between the two terms. The education transition matrix makes this concrete: 70.0\% of districts change educational category between the first (2015--2018) and second (2019--2022) terms.

We are careful about what this does and does not buy us. The ban makes \emph{turnover} institutionally guaranteed; it does not make the \emph{attributes of the incoming mayor} exogenous. Which mayor a district elects in 2018 remains a choice of that district, and it is precisely that choice which our negative controls interrogate.

\begin{table}[htbp]
\centering
\small
\begin{adjustbox}{max width=\textwidth}
\begin{tabular}{lcccc}
\toprule
\multicolumn{5}{l}{\textit{Panel A. Within-district variation}}\\
\midrule
\textbf{Variable} & \textbf{SD total} & \textbf{SD within} & \textbf{SD between} & \textbf{Within variance share}\\
\midrule
\exfp{} & 3.040 & 2.330 & 1.953 & 0.587\\
\eduf{}  & 3.722 & 2.525 & 2.736 & 0.460\\
\bottomrule
\end{tabular}
\end{adjustbox}

\vspace{6pt}
\begin{adjustbox}{max width=\textwidth}
\begin{tabular}{lcccc}
\multicolumn{5}{l}{\textit{Panel B. Education-category transition, term 1 (rows) $\rightarrow$ term 2 (columns), row-normalised}}\\
\toprule
\textbf{Term 1 $\backslash$ Term 2} & $\leq 11$ & 12--14 & 15--16 & $17+$\\
\midrule
$\leq 11$ & 0.444 & 0.186 & 0.216 & 0.154\\
12--14    & 0.377 & 0.204 & 0.233 & 0.186\\
15--16    & 0.311 & 0.137 & 0.335 & 0.217\\
$17+$     & 0.386 & 0.137 & 0.261 & 0.217\\
\bottomrule
\end{tabular}
\end{adjustbox}

\caption{Within-district variation and cross-term transition of mayoral human capital}
\label{tab:variation}
\floatnotes{\textit{Notes:} $N=13{,}136$ observations across 1{,}642 district municipalities, 2015--2022. Within variance share $=\text{SD within}^2/\text{SD total}^2$. Mayoral attributes are fixed at election and constant within each four-year term, so within variation arises from a single step at the 2019 transition. The off-diagonal mass (70.0\% of districts) reflects educational-category change across mayoral terms, induced by the no-immediate-re-election rule \citep{ley30305}. \textit{Source:} Authors' own work, based on JNE and MEF data.}
\end{table}

\subsection*{The conventional result: experience is consistent, education is fragile}

\begin{figure}[htbp]
\centering
\includegraphics[width=\textwidth]{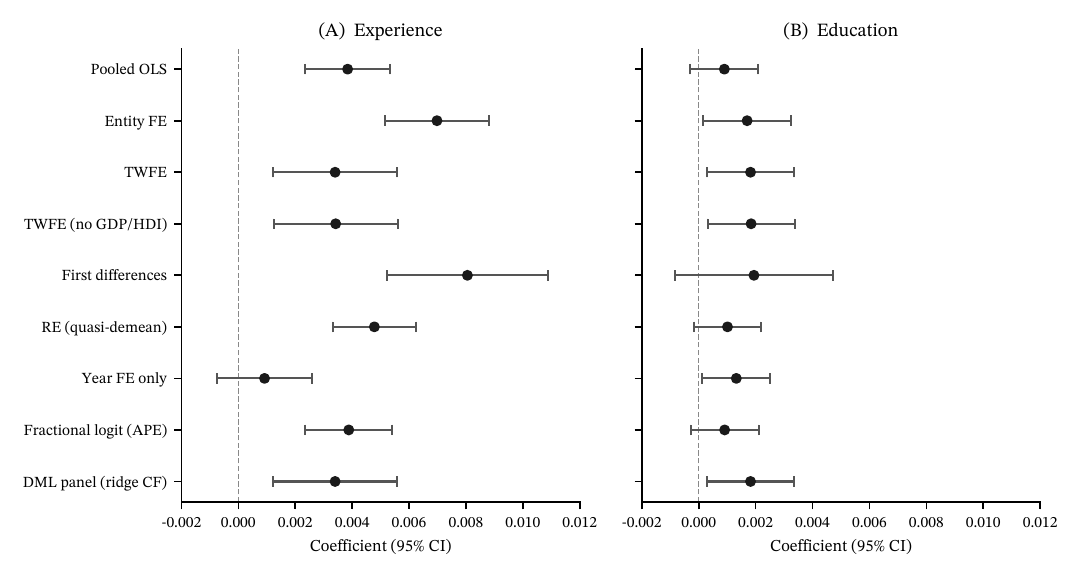}
\caption{Robustness of the human-capital effects across panel estimators}
\label{fig:robust}
\vspace{2pt}
\floatnotes{\textit{Notes:} Coefficient plot of the rows in Table~\ref{tab:main}, in the same order. The ridge DML row is a functional-form check rather than independent evidence; the effective battery consists of the remaining estimators. Panel~(A): experience. Panel~(B): education. Horizontal spikes are 95\% confidence intervals from district-clustered standard errors. The dashed vertical line marks zero. \textit{Source:} Authors' own work.}
\end{figure}

Prior public-management experience is positive and statistically significant under the main within and double-machine-learning specifications, ranging from 0.0033--0.0034 (TWFE/DML) to about 0.008 (first differences) in the linear models (Table~\ref{tab:main}; Figure~\ref{fig:robust} plots the coefficient battery across estimators). Formal education attains significance mainly in the entity fixed-effects and TWFE specifications ($\beta\approx0.0018$). The Hausman test ($H=130.55$; $p<0.001$) rejects random effects. The panel-DML estimator, which orthogonalises the fixed effects before district-clustered cross-fitting, yields an ATE for experience of 0.0033 (95\% CI $[0.0015, 0.0051]$), tightly coherent with the parametric within estimates. Fractional-logit average partial effects reinforce the pattern on the experience margin. Estimates are stable to the inclusion of investment-budget-quartile and transfer-intensity categories, to the exclusion of the university and institute indicators, and to Driscoll--Kraay and two-way clustering; they are also essentially unchanged when log district GDP and district HDI are dropped from the control set (TWFE experience $0.00340$ versus $0.00342$), so the within association does not operate solely through those measured district conditions. Education, already marginal, loses significance under region-year effects. Year fixed effects alone (without district fixed effects) drive the experience coefficient toward zero and insignificance ($0.00092$, SE $0.00085$). Year dummies do not remove between-district variation; that drop from pooled OLS is therefore informative about period or cohort confounding---including the 2019 mayoral turnover---rather than about time-invariant district differences.

Two design notes discipline the battery. First, first differences are about 2.4 times the TWFE estimate ($0.00805/0.00340$); under homogeneous effects and strict exogeneity the two should coincide. A district-cluster bootstrap of the difference has 95\% CI $[0.0014, 0.0067]$, so the gap is not sampling noise: it is a symptom of serial correlation, dynamics, or term-coincident shocks---and first differences in this design load heavily on the single 2018--2019 cohort step. Second, when the sample is restricted to 2019--2022---the only term in which immediate re-election was banned and no mayor is an incumbent from the previous term---a within comparison is impossible because the mayor is fixed. The district-level (between) association of experience with execution is then $0.0002$ (SE $0.0018$), essentially null. On the homologous Year-FE-only specification the window gives $0.00112$ (SE $0.00175$), against $0.00024$ (SE $0.00097$) in 2015--2018 and $0.00092$ (SE $0.00085$) on the full panel. A stacked, fully interacted regression with district-clustered errors cannot reject equality of the two term-specific coefficients (difference $0.00087$, SE $0.00199$; Wald $=0.19$, $p=0.66$), nor equality of the 2019--2022 window with the full panel (difference $0.00020$, SE $0.00155$; Wald $=0.02$, $p=0.90$): there is no attenuation on a homologous estimand, and if anything the post-ban coefficient is larger. Comparing a between estimand with the full-panel TWFE ($0.00340$) mixes a between with a within estimand. We cannot report the exact re-elected share in 2015--2018 without candidate-level JNE identifiers beyond the analysis file.

A Mundlak decomposition---year fixed effects plus district means---recovers the same within coefficient on experience as TWFE ($0.0034$, SE $0.0011$) and a negative coefficient on the district mean ($-0.0047$, SE $0.0018$). The implied between slope is about $-0.0013$: districts that usually elect more-experienced mayors do not execute more once the within step is separated.

\begin{center}
\begin{tabular}{lc}
\toprule
\textbf{Component} & \textbf{Coefficient (SE)}\\
\midrule
Within coefficient & 0.0034 (0.0011)\\
District-mean coefficient & $-0.0047$ (0.0018)\\
Implied between slope & $-0.0013$\\
\bottomrule
\end{tabular}
\end{center}

The implied between slope is derived from the sum of the within and district-mean coefficients; its standard error is not reported because it requires the covariance between those two estimated coefficients. Within-district AR(1) in execution is $-0.10$ (SE $0.01$), year-to-year mean reversion consistent with PIM revisions and with first differences exceeding TWFE.

The 2019 experience step is not an extrapolation of a pre-existing execution path. In 2015--2018, $\Delta$experience does not predict contemporaneous execution (the 2018 between coefficient is the largest in absolute value: $-0.0018$, $t=-1.24$). A district-fixed-effects interaction of $\Delta$experience with a linear year trend is $-0.0005$ ($t=-0.81$); event-study coefficients for 2016, 2017 and 2018, with 2015 as the base, are all $|t|<0.82$.

A third check addresses the 18\% mass at zero experience. A two-part reading of the within design is informative. On the intensive margin---district--years with strictly positive experience---the TWFE coefficient on years of experience remains positive and significant but is somewhat smaller than the full-sample linear coefficient ($0.00251$, SE $0.00117$, versus $0.00340$, SE $0.00111$; $N=10{,}667$). On the extensive margin, an indicator for any prior public-management experience is not significantly associated with execution once the same controls and two-way demeaning are applied ($0.00895$, SE $0.00912$, $t=0.98$). The association with experience is therefore not an artefact of the extensive margin alone: additional years among those with some experience still move with execution, while merely crossing from zero to positive experience does not, on its own, deliver a precise within estimate. On a budget-credibility margin, TWFE experience is associated with a lower probability of a failed execution year (execution below 50\%: $\beta=-0.0054$, $t=-2.67$).

Agreement across linear within estimators is not strong evidence of a causal leadership effect. Multiple estimators on the same data are correlated transformations; they establish robustness to \emph{functional form}, not to \emph{confounding}. On that narrower reading, experience behaves as expected under H2 in the full panel, while education is only weakly and conditionally consistent with H1. Whether either association survives selection is the subject of the diagnostics below.

\begin{table}[htbp]
\centering
\small
\begin{adjustbox}{max width=\textwidth}
\begin{tabular}{lcccccc}
\toprule
\textbf{Estimator} & Exp.\ $\beta$ & SE & Edu.\ $\beta$ & SE & $N$ & Clusters\\
\midrule
Pooled OLS & 0.00384 & 0.00076 & 0.00090 & 0.00061 & 13{,}025 & 1{,}642\\
Entity FE & 0.00698 & 0.00094 & 0.00170 & 0.00079 & 13{,}025 & 1{,}642\\
TWFE & 0.00340 & 0.00111 & 0.00182 & 0.00078 & 13{,}025 & 1{,}642\\
TWFE (no GDP/HDI) & 0.00342 & 0.00111 & 0.00184 & 0.00078 & 13{,}025 & 1{,}642\\
First differences & 0.00805 & 0.00145 & 0.00194 & 0.00141 & 11{,}383 & 1{,}642\\
RE (quasi-demean) & 0.00478 & 0.00075 & 0.00101 & 0.00060 & 13{,}025 & 1{,}642\\
Year FE only & 0.00092 & 0.00085 & 0.00132 & 0.00061 & 13{,}025 & 1{,}642\\
Fractional logit (APE) & 0.00388 & 0.00078 & 0.00091 & 0.00061 & 13{,}025 & 1{,}642\\
DML panel (ridge CF) & 0.00340 & 0.00111 & 0.00182 & 0.00078 & 13{,}025 & 1{,}642\\
\bottomrule
\end{tabular}
\end{adjustbox}
\caption{Main results: mayoral human capital and investment-budget execution}
\label{tab:main}
\floatnotes{\textit{Notes:} Outcome is the investment execution ratio. District-clustered robust standard errors. Controls (except the ``no GDP/HDI'' row) include age, ethical-record indicator, education and experience jointly, log district GDP and district HDI where indicated. Fractional logit reports average partial effects; DML uses two-way demeaning then ridge cross-fitting at the district level. $N$ is smaller than 13{,}136 where experience is missing. The headline DML ATE of 0.0033 $[0.0015, 0.0051]$ is the LassoCV specification reported in the text; the ridge check is 0.00340. At five decimal places the experience coefficients are $0.00340$, $0.00342$ and $0.00340$ and the education coefficients are $0.00182$, $0.00184$ and $0.00182$ for TWFE, TWFE (no GDP/HDI) and ridge DML, respectively. The ridge DML row is not presented as independent evidence: after two-way demeaning and with linear nuisance functions, it recovers the same within linear projection as TWFE. It is retained as a functional-form check; the effective independent estimators are the remaining rows of the battery. \textit{Source:} Authors' own work.}
\end{table}

\subsection*{COVID-19 and the limits of differential inference}

\begin{figure}[htbp]
\centering
\includegraphics[width=0.86\textwidth]{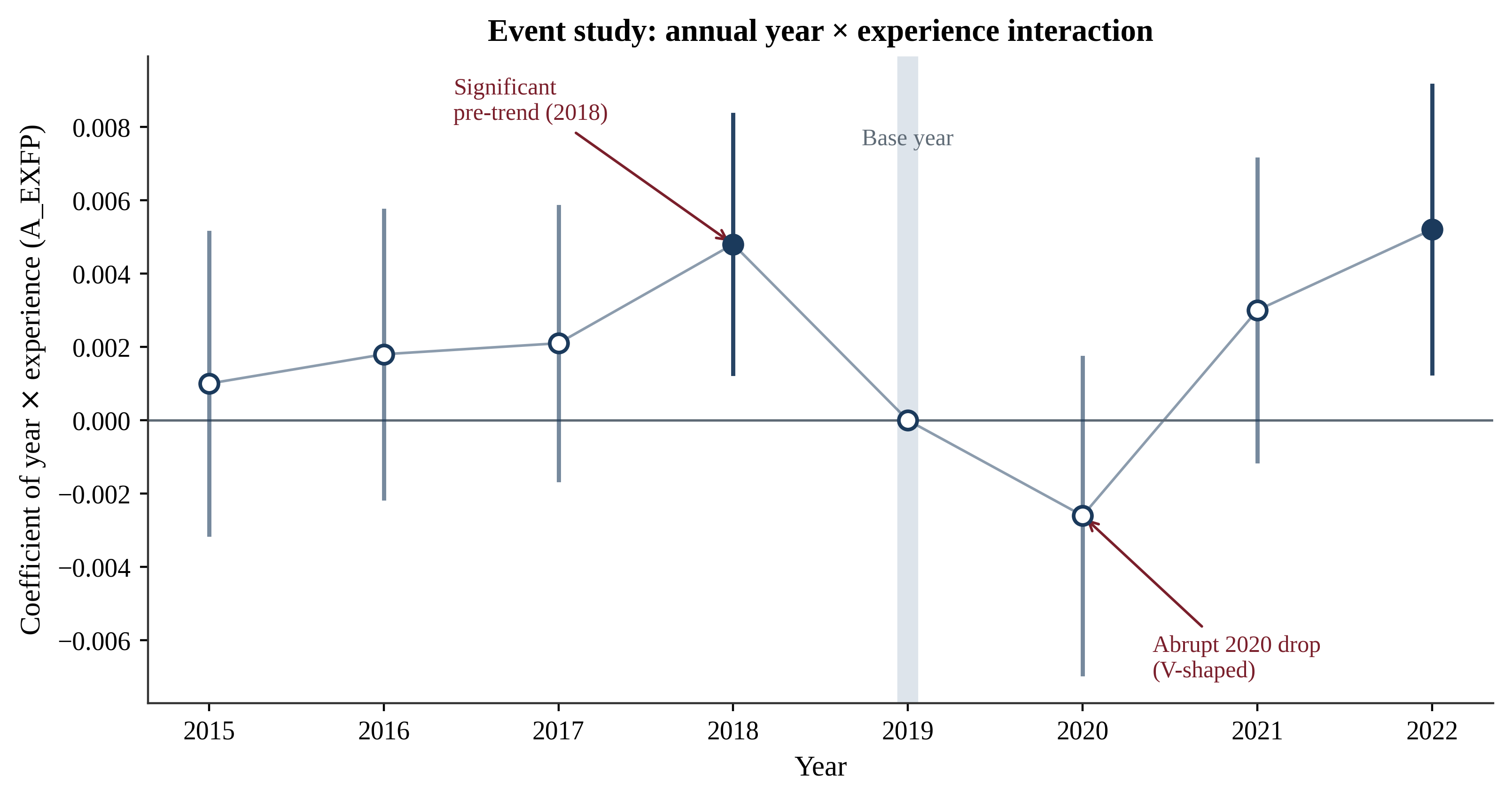}
\caption{Event study: annual coefficient of the year $\times$ experience interaction}
\label{fig:event}
\vspace{2pt}
\floatnotes{\textit{Notes:} annual coefficients with entity fixed effects and HC3 errors; 2019 is the base category. The significant 2018 coefficient (pre-treatment) and the V-shaped trajectory around 2020 evidence the violation of the pre-trends assumption and caution against a causal interpretation of the conventional event study. \textit{Source:} Authors' own work.}
\end{figure}

H3 asked whether the crisis revalued experiential knowledge. We treat the claim as exploratory and keep the exposition short. A conventional event study can look like amplification of the return to experience, but pre-trends are rejected ($\chi^2=13.74$; $\mathrm{df}=4$; $p=0.008$), and conditioning the rest of the analysis on that pretest would itself distort inference \citep{roth2022aeri}. The 2020 path is V-shaped rather than monotone, and within the second term alone---where the mayor is fixed---the post-2020 $\times$ experience interaction is essentially zero ($\beta=0.0002$, SE $=0.0030$). Cohort turnover in 2019, not the shock itself, drives most of the apparent recomposition; staggered-adoption estimators disagree in sign because they are mismatched to a common-timing, continuous-treatment design. The interaction is the one hypothesis in the family that fails multiple-comparison correction. H3 is not supported as a causal claim (Figure~\ref{fig:event}).

\subsection*{Heterogeneity}

The average associations leave open whether the experience gradient is uniform. On the source zone label, TWFE experience is larger in urban municipalities ($\beta=0.0048$; $t=2.46$) than in rural ones ($\beta=0.0027$; $t=1.99$). Education is only marginally associated with execution and is not robust across zone. The education $\times$ urban interaction is not significant ($p=0.79$). The honest causal forest yields a predominantly positive distribution of conditional effects (mean CATE $=0.0051$; the discrepancy with the DML ATE of 0.0033 reflects distinct estimands---an unweighted mean of CATEs versus an orthogonalised average effect---not a contradiction). The dominant moderators are district GDP (relative importance 36\%), the pandemic period (25\%), and district HDI (17\%); mayoral education contributes 6\% relative importance. The machine learning points at the district, not at the mayor.

\subsection*{Selection: negative controls, sensitivity, and partial identification}

H4 asks which part of the human-capital--execution association is selection. Both lagged district HDI and district GDP---outcomes the mayor cannot have caused within their term---are significantly predicted by mayoral human capital (education $\rightarrow$ lagged HDI: $\beta=0.0033$, $p<0.001$; experience $\rightarrow$ GDP: $\beta=0.0625$, $p<0.001$). To rule out short-run mayoral influence, we repeat the exercise with strictly pre-determined outcomes measured before 2015. Table~\ref{tab:nc} reports the pre-2015 specifications.

The between negative control~\eqref{eq:nc-between} is unambiguous: district-mean experience predicts 2015 log GDP ($0.0787$, SE $0.0230$, $t=3.43$). That is the contamination district fixed effects remove by construction, and it is enough to unsettle any cross-sectional reading of the mayor's CV. The specification aligned with FE/TWFE/DML is the first difference~\eqref{eq:nc-fd}. Inter-term $\Delta$experience does not predict 2015 log GDP ($-0.0077$, SE $0.0110$, $t=-0.69$). The null is not an artefact of the 49.9\% of districts with a zero step: among the 823 changers the coefficient is $0.0087$ (SE $0.0124$, $t=0.70$); among the 763 with $|\Delta S|\ge 2$ it is $0.0028$ (SE $0.0125$, $t=0.22$). Approximate 80\% minimum detectable effects on those standard errors are $0.031$--$0.035$ on the log-GDP scale---well below the between-district selection coefficient of $0.0787$, so a within-aligned selection effect of the magnitude documented between districts would be detected and the null first-difference result is not an artefact of low power. The same $\Delta S$ \emph{does} predict the change in execution (district-level first difference $0.0033$, SE $0.0011$, $t=2.99$; changers $0.0025$, SE $0.0012$, $t=2.04$). Levels specification~\eqref{eq:nc-within} still loads on both terms, as expected when selection is between districts. First-difference HDI is negative ($-0.0025$, SE $0.0009$, $t=-2.85$): districts that \emph{raise} experience in 2019 were slightly less developed in 2015. H4 is therefore not ``no within selection of any kind.'' It is that the GDP-aligned falsification the within design requires is passed, while between-district selection is not. The between GDP control is not uniform. Split by 2015 HDI tercile, mean experience does not predict 2015 log GDP in the poorest third ($-0.0147$, SE $0.0243$, $t=-0.60$, $N=547$) and does in the richest third ($0.1509$, SE $0.0424$, $t=3.56$, $N=547$). Documented selection is a property of already-developed districts.

Because the negative-control outcome and the execution outcome are on different scales, raw coefficients cannot be compared directly. Standardising the \emph{between} experience coefficient with the standard deviation of district-mean experience ($1.920$) and of 2015 log GDP ($1.615$) yields $0.094$ SD (district-cluster bootstrap 95\% CI $[0.045, 0.148]$). The TWFE execution association, standardised with the observation-level SDs of experience ($3.035$) and execution ($0.222$), is $0.046$ SD (district-cluster bootstrap 95\% CI $[0.017, 0.076]$).\footnote{An earlier draft compared the raw contemporaneous GDP coefficient ($0.0625$) with the raw execution coefficient ($0.0033$) and reported a ratio of roughly nineteen. That comparison is invalid. A further pitfall is standardising the between negative control with the observation-level SD of experience ($\approx3.04$) rather than the SD of the district mean, or treating the raw second-term coefficient $0.0891$ (Panel~B) as if it were already a standardised between association. The experience--GDP coefficients are four distinct objects: $0.0625$ is the contemporaneous within-term raw association, $0.0787$ is the pre-2015 between (district-mean) raw association, $0.0891$ is the second-term level raw association, and $0.094$ is the standardised between association ($0.0787\times1.920/1.615$).} The $0.094$ versus $0.046$ comparison is therefore a comparison of standardised associations on different outcomes, not an estimate of the share of $\hat\beta_S$ that is selection. Under the proportionality assumption in Section~\ref{sec:method} the negative control remains a diagnostic of selection, not a debiased ATE. A paired district-cluster bootstrap of the difference (resampling districts once and re-estimating both standardised associations on the same draw) gives 95\% CI $[-0.013, 0.105]$ with 92.5\% of draws positive, so the two associations are not statistically distinguishable; the comparison is a diagnostic of the pattern of selection, not a test of the share of selection.

\begin{table}[htbp]
\centering
\small
\begin{adjustbox}{max width=\textwidth}
\begin{tabular}{llcccc}
\toprule
\textbf{Specification} & \textbf{Regressor} & $\beta$ & SE & $t$ & $N$\\
\midrule
\multicolumn{6}{l}{\textit{A. Between: 2015 log GDP $\sim$ district-mean human capital + region FE}}\\
 & Mean experience & 0.0787 & 0.0230 & 3.43 & 1{,}642\\
 & Mean education & 0.1216 & 0.0135 & 9.02 & 1{,}642\\
\midrule
\multicolumn{6}{l}{\textit{B. Levels (eq.~\ref{eq:nc-within}): 2015 log GDP $\sim$ term-2 and term-1 experience + region FE}}\\
 & Second-term experience & 0.0891 & 0.0259 & 3.44 & 1{,}642\\
 & First-term experience & 0.0334 & 0.0128 & 2.61 & 1{,}642\\
\midrule
\multicolumn{6}{l}{\textit{C. First differences (eq.~\ref{eq:nc-fd}): 2015 log GDP $\sim$ $\Delta$ human capital + region FE}}\\
 & $\Delta$ experience (all districts) & $-0.0077$ & 0.0110 & $-0.69$ & 1{,}642\\
 & $\Delta$ experience (changers only) & 0.0087 & 0.0124 & 0.70 & 823\\
 & $\Delta$ education (all districts) & 0.0091 & 0.0071 & 1.28 & 1{,}642\\
\midrule
\multicolumn{6}{l}{\textit{D. First differences: 2015 HDI $\sim$ $\Delta$ human capital + region FE}}\\
 & $\Delta$ experience & $-0.0025$ & 0.0009 & $-2.85$ & 1{,}642\\
\bottomrule
\end{tabular}
\end{adjustbox}
\caption{Negative-outcome controls: pre-2015 district conditions and mayoral human capital}
\label{tab:nc}
\floatnotes{\textit{Notes:} One observation per district. Heteroskedasticity-robust standard errors (the district identifier is the unit of observation, so clustering by district does not add within-cluster information). All specifications include region fixed effects. Standardising Panel~A mean experience with $\mathrm{SD}(\bar S_d)=1.920$ and $\mathrm{SD}(\ln\mathrm{GDP}_{2015})=1.615$ yields $0.094$ SD (bootstrap 95\% CI $[0.045, 0.148]$). The TWFE execution association standardised with observation-level $\mathrm{SD}(S)=3.035$ and $\mathrm{SD}(Y)=0.222$ is $0.046$ SD. About 49.9\% of districts have $\Delta$experience $=0$; the changers-only row restricts to the rest. Approximate 80\% MDEs on the Panel~C experience SEs are $0.031$ (all) and $0.035$ (changers). The raw $0.0891$ in Panel~B and the standardised $0.094$ from Panel~A are different objects. A paired district-cluster bootstrap of the difference between the standardised between selection association (0.094 SD) and the standardised execution association (0.046 SD) gives a 95\% CI of $[-0.013, 0.105]$; the two associations are not statistically distinguishable. \textit{Source:} Authors' own work.}
\end{table}

\begin{figure}[htbp]
\centering
\includegraphics[width=\textwidth]{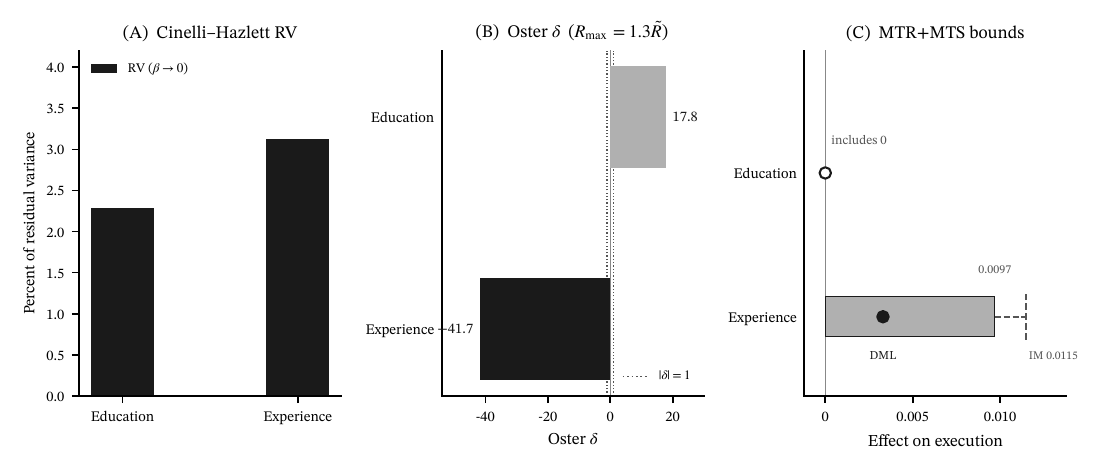}
\caption{Sensitivity and partial-identification profile for education and experience}
\label{fig:sens}
\vspace{2pt}
\floatnotes{\textit{Notes:} Panel~(A): Cinelli--Hazlett robustness values (percentage of residual variance a confounder must explain to reduce the effect to zero). We do not treat any fixed percentage as a conventional cutoff \citep{cinelli2020}. Panel~(B): Oster's $\delta$ on the TWFE residualisation. Dotted guides mark $|\delta|=1$. Education $17.8$; experience $-41.7$ (adding covariates raises the experience coefficient). An earlier residualisation produced $-0.39$ and $0.18$. Panel~(C): Manski--Pepper identified set under MTR+MTS. Education includes 0 and has no reconstructable upper bound in the analysis file. For experience the set is $[0, 0.0097]$; the dashed cap is the Imbens--Manski-style 95\% interval for that set ($0.0115$). The filled marker is the DML point estimate ($0.0033$). The zero lower endpoint is implied by MTR. \textit{Source:} Authors' own work.}
\end{figure}

Sensitivity analysis (Figure~\ref{fig:sens} and Table~\ref{tab:sens}) does not reverse that partition. Cinelli--Hazlett robustness values on the TWFE model indicate that a confounder explaining about 3.1\% of residual variance would nullify the experience coefficient (2.3\% for education). Those values are small; \citet{cinelli2020} do not prescribe a 3\% cutoff, and we do not treat one as a finding. Benchmarking against log district GDP after the within transformation leaves the point estimate essentially unchanged at 1$\times$ and 3$\times$ relative strength. That is a statement about the \emph{benchmark}, not about robustness: after two-way demeaning, GDP is almost orthogonal to the 2019 experience step (a confounder on the order of 170 times within-GDP would be required to zero the estimate). Mayor age is the least mute observed benchmark (about 18 times). Age serves as a benchmark not because it is a capacity proxy but because it is an observed mayor covariate that retains within-district variation after demeaning; the natural capacity proxies (log district GDP, district HDI) are almost orthogonal to the 2019 step and so provide no usable benchmark. Even a confounder three times as strong as age leaves a positive point estimate. Because direct bureaucratic-capacity measures are not in the analysis file, mayor age is the strongest observed benchmark available; this does not make age a capacity proxy. The small robustness values and the near-orthogonality of the observed benchmarks together deliver one message: a confounder strong enough to nullify the experience coefficient would have to be far stronger than any covariate we observe, but benchmarking does not by itself prove that no such confounder exists. Oster's $\delta$ computed on the TWFE residualisation is large and negative for experience ($-41.7$), because adding observed covariates slightly \emph{raises} the coefficient---the opposite of classical attenuation. Education's $\delta$ on the same residualisation is $17.8$: observables attenuate that coefficient, but not below Oster's $|\delta|>1$ threshold. (An earlier draft reported $0.18$ and $-0.39$ on a different residualisation.) Those Oster magnitudes do not overturn the small robustness values. E-values are omitted: they target risk ratios, whereas the robustness value is the appropriate metric for a continuous, bounded outcome \citep{vanderweele2017}.

Having discarded the instrumental strategy, we report Manski--Pepper bounds under monotone treatment response (MTR) and monotone treatment selection (MTS). The lower endpoint is 0 by MTR---mechanical, not a discovery. The reconstructable upper bound is the median-split slope of execution on experience, $0.0097$ (district-cluster bootstrap 95\% CI $[0.0077, 0.0118]$). An Imbens--Manski-style 95\% interval for the identified \emph{set} $[0,\hat U]$ is $[0, 0.0115]$ \citep{imbensmanski2004}. An earlier draft reported an upper endpoint of $0.0155$ that we cannot reconstruct from the analysis file; we replace it. Data plus MTR+MTS still include a null causal effect.

Correcting the family of heterogeneity and exploratory interaction tests via Benjamini--Hochberg and Romano--Wolf, eight of nine hypotheses survive false-discovery-rate control at 5\%. The exception is the COVID $\times$ experience interaction. We do not apply that multiple-comparison correction to the main experience and education coefficients in Table~\ref{tab:main}: those are the primary, pre-specified associations (H1--H2), not a search over many moderators. H4 is supported in the form restated above: between-district selection on pre-determined GDP is detectable; the first-difference control aligned with the within step is not, including among changers. That is not a causal certificate for the within association.

\begin{table}[htbp]
\centering
\small
\begin{adjustbox}{max width=\textwidth}
\begin{tabular}{lcc}
\toprule
\textbf{Test} & \textbf{\eduf{}} & \textbf{\exfp{}}\\
\midrule
Robustness value RV (reduce $\beta \to 0$)        & 2.29\%           & 3.13\%\\
Oster's $\delta$ ($R_{max}=1.3\tilde{R}$)          & $17.8$           & $-41.7$\\
Manski--Pepper bounds (MTR + MTS)                  & includes 0       & $[0, 0.0097]$\\
Panel-DML ATE [95\% CI]                            & n/a              & 0.0033 $[0.0015, 0.0051]$\\
Negative control (pre-2015, standardised)          & 0.073 SD (HDI)   & 0.094 SD (log GDP) $[0.045, 0.148]$\\
FDR-adjusted (primary coefficients)                & no               & no\\
\bottomrule
\end{tabular}
\end{adjustbox}

\caption{Sensitivity and partial-identification results for mayoral education and experience}
\label{tab:sens}
\floatnotes{\textit{Notes:} Robustness values and Oster's $\delta$ are computed on the two-way-demeaned model ($N=13{,}025$ district--years; 1{,}642 clusters). Experience $\delta=-41.7$; education $\delta=17.8$. An earlier residualisation produced $0.18$ and $-0.39$. The balanced panel has 13{,}136 observations; $N$ falls where experience is missing. Negative-control coefficients use the between (district-level) specification ($N=1{,}642$ districts) and are reported in standard-deviation units of their respective outcomes ($\beta\cdot\text{SD}(X)/\text{SD}(Y)$), with $\mathrm{SD}(X)$ equal to the SD of district-mean experience. For reference, the standardised execution association is $0.046$ SD (95\% CI $[0.017, 0.076]$) for experience and $0.030$ SD for education. A paired district-cluster bootstrap of the difference between the standardised between selection association and the standardised execution association gives a 95\% CI of $[-0.013, 0.105]$; the two associations are not statistically distinguishable. FDR is applied to the heterogeneity and exploratory-interaction family, not to these primary H1--H2 coefficients (see text). Robustness values are small; we interpret them via covariate benchmarking \citep{cinelli2020}, not a fixed percentage threshold. MTR $=$ monotone treatment response; MTS $=$ monotone treatment selection. The zero Manski--Pepper lower bound is imposed by MTR. The reconstructable upper bound is the median-split slope $0.0097$; an Imbens--Manski-style 95\% interval for the set is $[0, 0.0115]$.}
\end{table}

\subsection*{External validity}

\begin{figure}[htbp]
\centering
\includegraphics[width=0.92\textwidth]{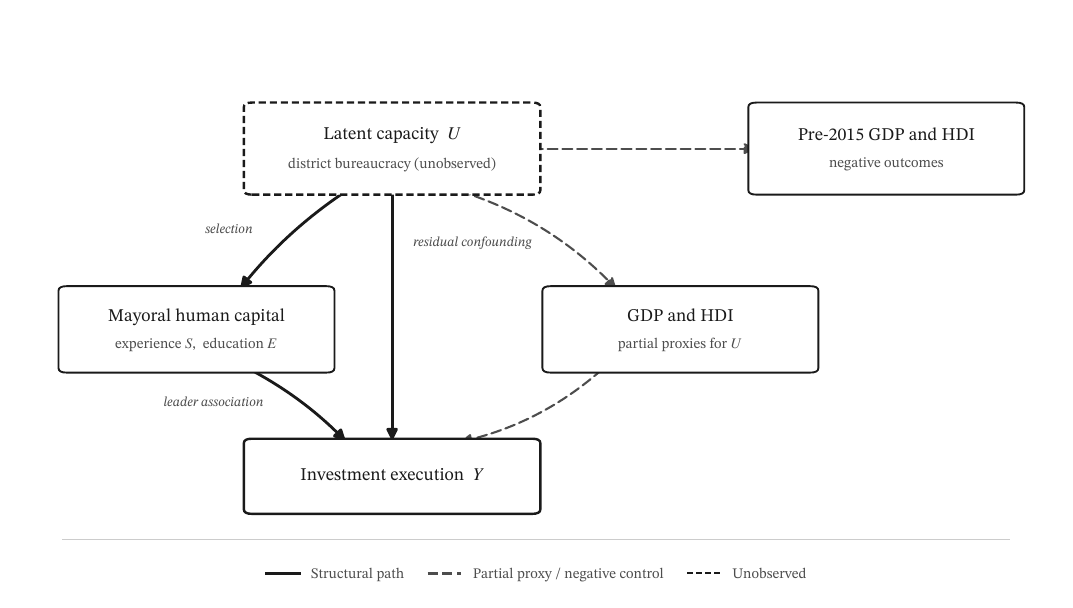}
\caption{Directed acyclic graph of mayoral human capital and investment-budget execution}
\label{fig:dag}
\vspace{2pt}
\floatnotes{\textit{Notes:} Solid arrows are the three structural paths. Latent bureaucratic capacity $U$ selects mayoral human capital and affects execution through a \emph{direct} channel that does not run through measured GDP/HDI---consistent with the near-invariance of the TWFE experience coefficient when those controls are dropped. Dashed arrows mark GDP and HDI as partial proxies for $U$, not as exhaustive confounders, and mark pre-2015 GDP/HDI as negative outcomes the incoming mayor cannot have caused. The dashed box denotes that $U$ is unobserved; district fixed effects remove only its time-invariant part. Distinctions are by line pattern, so the figure remains legible in greyscale. \textit{Source:} Authors' own work.}
\end{figure}

Whether these patterns travel is a separate question. A transportability analysis \citep{pearl2014} reweights the estimated association by the urban--rural composition of hypothetical target populations; the association with experience remains within a narrow band under substantial variation in territorial composition. That reweighting varies only the urban--rural mix, not the deeper structural dimensions---administrative capacity, budget size, transfer dependence---on which districts are heterogeneous, so the exercise is suggestive rather than a certified transportability claim. In particular, it does not imply generalisation to other countries---for example Colombia, Mexico, Indonesia or the Philippines---whose territorial organisation, transfer systems and mayoral selection rules differ from Peru's. Figure~\ref{fig:dag} summarises the directed acyclic graph that structures the design and makes the selection channel explicit.

\section{Discussion}
\label{sec:discussion}

\subsection*{What we found, and what it costs the leadership literature}

On a national panel and with a wider battery of estimators than is customary, we recover the finding that anchors a large literature: mayors with prior public-management experience preside over better budget execution. Between districts, the same human capital predicts conditions measured before 2015. That is enough to unsettle the cross-sectional literature \citep{besley2011,avellaneda2009,freier2016,wan2021}. Within districts, the 2019 experience step still moves execution and does not predict 2015 log GDP---including among districts that change experience. The aligned falsification is passed on GDP; it is not a licence to read the within coefficient as a causal leader effect. Asked what moderates the association, the honest causal forest returns district GDP, district HDI, and the pandemic; mayoral education contributes 6\% relative importance.

The comparison with \citet{klindt2026} is useful precisely because the two papers disagree about something. They find that the between-municipality partisanship effect vanishes once structural controls are added, and read this as evidence that structure \emph{accounts for} between-variation better than partisanship. We would put it more sharply: where structure also selects the leader, controlling for structure does not merely absorb a rival explanation; it partially absorbs the leader variable itself, because the leader is downstream of structure. Their within-municipality interaction survives because it exploits \emph{change} in leadership against \emph{change} in conditions---the design closest to breaking the selection link. Together the studies suggest that the durable ``leader effect'' in local government may be smaller than either the old or the new view has assumed, and that what remains is conditional, short-run, and visible mainly in within-unit designs.

We do not claim that mayors are irrelevant. The within point estimate is positive; the aligned GDP negative control does not detect selection on the 2019 step; under MTR the Manski--Pepper lower bound is mechanically non-negative while the identified set still includes zero; and classical measurement error in self-reported CVs would bias coefficients towards zero. The claim is narrower: cross-sectional estimates do not separate the mayor from the municipality, and the within association, though it survives that particular falsification, is small and not certified as causal.

\subsection*{Even taken at face value, the effect is small}

Even if one set selection aside, the magnitude would still discipline the rhetoric. An additional year of mayoral experience is associated with 0.0033 to 0.0069 additional units of execution on a $[0, 1]$ scale, against an outcome standard deviation of 0.222. Moving a mayor from zero experience to the sample mean of about 2.97 years is associated with roughly one percentage point of execution---under one-twentieth of a standard deviation. Absolute soles translations require PIM levels, which are not in the public analysis file; for an illustrative modified investment budget of 10 million soles, a one-year increase would correspond to roughly 33{,}000 to 69{,}000 soles of additional executed investment per year---not nothing in a capacity-constrained municipality, but far from a transformative leadership effect. Policy recommendations organised around candidate credentials have rested on an association that is both small and, as the selection analysis shows, not securely causal.

The pandemic result does not reverse that assessment. Design scrutiny rules out a causal reading of any crisis amplification of experience: parallel trends fail, the second-term-only interaction is null, and the interaction does not survive multiple-comparison correction \citep{roth2023,callaway2024}. We report it as an exploratory pattern, not a finding.

\subsection*{Implications for practice and method}

The policy conversation should move from the person to the apparatus---a partial vindication of \citet{peterson1981} in a setting he never considered. Because the evidence for credential effects is small, and because the between-district association is selection-contaminated, raising educational requirements for candidates is not a defensible lever for improving execution. District characteristics dominate the heterogeneity analysis and predict mayoral attributes themselves, so the more plausible binding constraint is the municipal bureaucracy: merit-oriented subnational civil-service systems \citep{servir2021}, stable technical teams, project-bank capacity, and continuity of investment-management routines. That recommendation follows the \emph{logic} of the findings; it is not itself causally identified here. Capacity-building for incoming mayors, as distinct from credential screening of candidates, remains reasonable on the modest evidence that experience, rather than education, is the more consistent correlate.

Methodologically, negative-outcome controls require no new data, no instrument and no design innovation---only the willingness to regress the treatment on something it cannot have caused and to report the answer. Observational studies of leader effects should adopt them as routinely as pre-trend tests after \citet{roth2023}, with the pretest caution of \citet{roth2022aeri}. The diagnostic is portable wherever leaders are selected by the units whose performance is measured.

\subsection*{Limitations}

Several limitations remain. Mayoral attributes are constant within term, so fixed-effects identification rests on a single step per district (Section~\ref{sec:method}); power is lower than the observation count suggests (a conservative minimum detectable effect for the within design, treating each district as contributing one treatment step, is $0.0058$, above the TWFE point estimate), and term-coincident shocks are a live threat. The execution ratio uses the modified budget (PIM) as denominator; PIA amounts are not in the analysis file. The university and institute counts used as capacity proxies are assigned at provincial scale (Table~\ref{tab:desc}). The low within $R^2$ indicates that mayoral human capital explains only a minor fraction of execution variation. Political alignment with the central or regional government and Municipal Council fragmentation are not in our analysis file. Direct measures of bureaucratic capacity (staffing, management instruments, project-bank certifications) are in principle available from public sources such as the Registro Nacional de Municipalidades (RENAMU, INEI) and MEF incentive and investment systems; we did not merge them here, and their absence very likely contributes to the selection the negative controls reveal. The absence of direct bureaucratic-capacity measures also limits sensitivity benchmarking: no observed capacity benchmark stronger than mayor age is available in the current analysis file. Pre-2015 investment-execution series from the same MEF source as the outcome---the most powerful negative control on a common metric---are likewise left for the next revision once district identifiers are linked.

Education and experience come from self-reported JNE curricula vitae and are subject to measurement error; if classical, this attenuates coefficients, and the ethical-record measure is an imperfect proxy for the non-cognitive skills \citet{heckman2000} emphasises. About 18\% of non-missing experience observations are exactly zero; the two-part check above shows that the intensive-margin years association remains positive while an extensive-margin indicator is imprecise. We did not implement a close-election regression-discontinuity design. Municipal electoral results for 2014, 2018 and 2022 are published by the Oficina Nacional de Procesos Electorales (ONPE); the binding obstacle is not the absence of digitised vote margins but the reliable linkage of candidate-level CVs (JNE) to electoral returns (ONPE). Completing that linkage---and merging RENAMU capacity measures and pre-2015 execution series---is the natural next step. Those three merges are not in this version; we list them as open data items rather than as a methodological justification. The exact share of re-elected incumbents in the 2015--2018 cohort cannot be computed without candidate-level JNE identifiers; the paper therefore treats the 2015--2018 cohort composition qualitatively.

\section{Conclusion}
\label{sec:conclusion}

This study examined whether the positive association between mayoral experience and municipal investment-budget execution reflects a leader effect or, instead, the capacity of municipalities to select more experienced leaders. Using a national panel of 1,642 Peruvian district municipalities from 2015 to 2022, we find that prior public-management experience is consistently associated with higher budget execution, while formal education shows a substantially weaker relationship.

The identification analysis reveals an important distinction. Across municipalities, mayoral human capital predicts GDP and HDI measured before the mayor took office, providing clear evidence that cross-sectional comparisons are affected by selection on pre-existing municipal conditions. Within municipalities, however, the change in mayoral experience that identifies the main estimates does not predict pre-determined district GDP, including among municipalities that actually change experience. The same change remains positively associated with budget execution. The within association therefore survives this negative-outcome falsification, although sensitivity analysis and partial-identification bounds do not support a clean causal interpretation.

These findings do not imply that political leadership is irrelevant. Rather, they show that the empirical meaning of a leader effect depends critically on the source of variation used to identify it. Cross-sectional estimates combine characteristics of the mayor with characteristics of the municipality that selects that mayor. Within-municipality designs remove an important part of this selection problem, but they do not eliminate all sources of confounding. This distinction also tempers the policy interpretation of mayoral credentials and shifts attention toward municipal administrative capacity, including the continuity of technical teams, professionalized public management, and institutionalized investment-management routines. These mechanisms remain hypotheses for future research rather than effects identified in this study.

The broader contribution is therefore methodological. Negative-outcome controls make visible a source of selection that conventional leader-effect estimates can easily conceal, while also showing when that selection does not survive the variation used for identification. For studies of political leadership, distinguishing the leader from the institution that selects the leader is not a secondary robustness exercise. It is part of the identification problem itself.

\section*{Acknowledgements}
The authors thank the anonymous reviewers for their valuable comments and suggestions.

\section*{Disclosure statement}
No potential conflict of interest was reported by the author(s).

\section*{Funding}
This research received no specific external funding.

\section*{Data availability statement}
Data derive entirely from public official sources (MEF \emph{Consulta Amigable}, JNE, INEI, UNDP, SUNEDU, MINEDU). The minimally processed dataset, the complete replication code reproducing every table and figure, a codebook, and a manifest with SHA-256 checksums and pinned software versions will be deposited in a public repository; an anonymised link (and DOI when assigned) is provided to the editors at submission, accessible to reviewers and not conditioned on acceptance.

\section*{Supplemental material}
This version does not attach a separate online appendix. Every estimate cited in the text is reported in the main tables and figures. A public replication deposit (data, code, codebook and checksums) is planned at submission; it is not yet assigned a DOI.

\begingroup
\raggedright

\endgroup

\end{document}